\documentclass{vgtc}                          

\ifpdf
  \pdfoutput=1\relax                   
  \usepackage{graphicx}                
  \DeclareGraphicsExtensions{.pdf,.png,.jpg,.jpeg} 
\else
  \usepackage{graphicx}                
  \DeclareGraphicsExtensions{.eps}     
\fi%

\graphicspath{{figures/}{pictures/}{images/}{./}} 

\usepackage{microtype}                 
\PassOptionsToPackage{warn}{textcomp}  
\usepackage{textcomp}                  
\usepackage{mathptmx}                  
\usepackage{times}                     
\usepackage{cite}                      
\usepackage{tabu}                      
\usepackage{booktabs}                  

\usepackage{xcolor}
\usepackage{appendix}
\usepackage{stfloats}

\onlineid{0}

\vgtccategory{Research}

\vgtcinsertpkg

\title{What are Data Insights to Professional Visualization Users?}

\author{
Po-Ming Law
\thanks{e-mail: pmlaw@gatech.edu}\\ %
\scriptsize Georgia Institute of Technology %
\and 
Alex Endert 
\thanks{e-mail: endert@gatech.edu}\\ %
\scriptsize Georgia Institute of Technology %
\and 
John Stasko
\thanks{e-mail: stasko@cc.gatech.edu}\\ %
\scriptsize Georgia Institute of Technology
}

\abstract{
While many visualization researchers have attempted to define data insights, little is known about how visualization users perceive them. We interviewed 23 professional users of end-user visualization platforms (e.g., Tableau and Power BI) about their experiences with data insights. We report on seven characteristics of data insights based on interviewees’ descriptions. Grounded in these characteristics, we propose practical implications for creating tools that aim to automatically communicate data insights to users.
} 

\CCScatlist{
  \CCScatTwelve{Human-centered computing}{Visu\-al\-iza\-tion}{Visu\-al\-iza\-tion theory, concepts and paradigms}{}
}

\begin{document}


\firstsection{Introduction}

\maketitle

The visualization community has recognized insight as a core purpose of visualizations~\cite{purpose}. While developing technologies that facilitate the process of gaining data insights, many researchers have articulated multiple definitions of insight. North~\cite{north} conceptualizes insights as complex, deep, qualitative, unexpected, and relevant revelations. Besides considering insight knowledge or information, Chang et al.~\cite{chang} believe that an insight can also be regarded as a moment of enlightenment. Despite the efforts to define data insights, little is known about how visualization users perceive them.

\begin{figure}[t!]
	\centering
	\includegraphics[width=\linewidth]{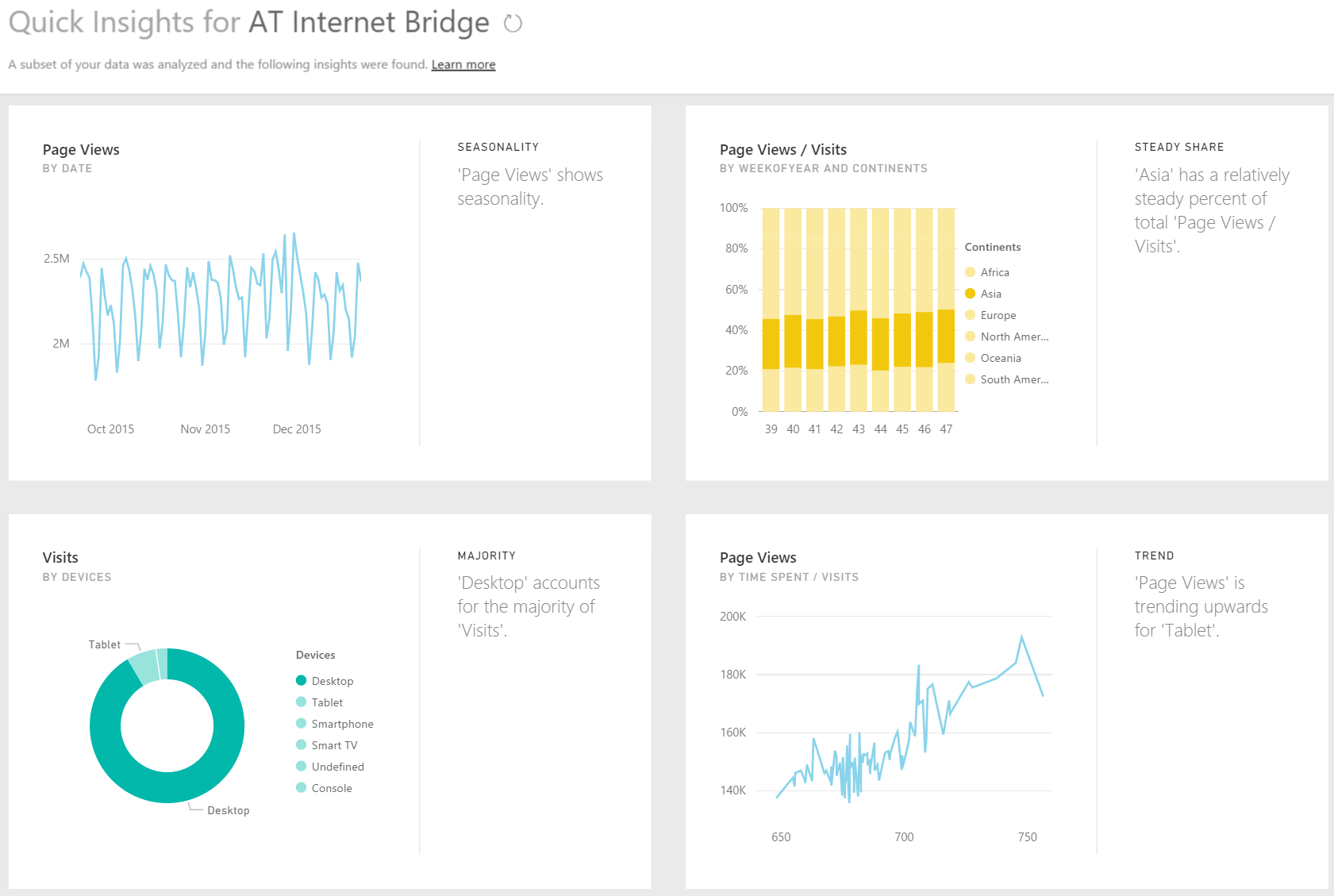}
	\vspace{-6mm}
	\caption{Many visualization systems have functionality that recommends potentially insightful visualizations and data facts~\cite{voder, datashot} to users. The figure shows Quick Insights in Power BI~\cite{quickInsights}.}
	\vspace{-5mm}
	\label{powerBI}
\end{figure}

Why care about visualization users’ perceptions of data insights? Understanding their perceptions could offer implications for designing tools that automatically generate data insights. Some researchers have envisioned automated systems (Fig.~\ref{powerBI}) that communicate data insights with similar qualities to those users glean through construction, manipulation, and interpretation of visualizations~\cite{newPaper}. These systems can accelerate knowledge discovery from data and lower the barrier to analysis for non-expert analysts. To create tools for automating data findings that are insightful to visualization users, we must understand what data insights are to these users.

Thus, we interviewed 23 practitioners who utilize end-user visualization platforms such as Tableau~\cite{tableau} and Power BI~\cite{powerBI} (hereafter, \textit{visualization users}) to investigate the characteristics of data findings they considered insightful. Being a large group of users who use visualizations to glean data insights as part of their work, these users can offer valuable perspectives on data insights and their automation.

From the data insights described by interviewees, we identified seven characteristics of data insights: actionable, collaboratively refined, unexpected, confirmatory, spontaneous, trustworthy, and interconnecting. Based on these characteristics, we propose implications for designing tools that aim to automatically generate data insights. Example design ideas include utilizing multiple information sources to generate more nuanced data insights, incorporating validation mechanisms to inspire user trust, and eliciting users’ expectations to provide more relevant data insights. We hope that the design implications we derive could provide guidance for designers and researchers when studying “automated insight” tools.

\section{Related Work}

\subsection{Prevailing Perspectives on Data Insights}

Broadly, visualization researchers have considered insight from three perspectives: a unit of information or knowledge, a psychological state, and an analysis by-product.

{The visualization community often defines an insight as a unit of information or knowledge~\cite{dik, insightEval}.} A widely-accepted definition is offered by North~\cite{north} who described insights as complex, deep, qualitative, unexpected, and relevant revelations. Chang et al.~\cite{chang} observed that visualization researchers often used “insight” in the same sense as “knowledge” or “information.” Chen et al.~\cite{insightMgmt} defined an insight as a fact that is evaluated through a mental model to inspire a psychological state of enlightenment. Researchers who employ insight-based evaluation methods~\cite{insightBased1} for visualization evaluation often define data insights as facts, observations, generalizations, and hypotheses from data (e.g.,~\cite{MCP}). In characterizing the insights quantified selfers gained from their data, Choe et al.~\cite{personalData} regarded data observations such as trends and comparisons as insights.

Insight can also be viewed as a psychological state. Laypeople often describe an insight as a eureka moment~\cite{eureka}, a light bulb moment~\cite{lightbulb}, and an aha moment~\cite{aha}. In cognitive science, a prevailing view is that an insight occurs when people transition from a state of not knowing how to solve a problem to a state of knowing how to solve it~\cite{cogSci}. Chang et al.~\cite{chang} called this type of insight a spontaneous insight and noted that insights as units of knowledge serve a knowledge-building function that promotes spontaneous insights.

From the first perspective, an insight is considered a finding---the end result of an analysis. A contrasting view is to regard an insight as a by-product of an analysis. Yi et al.~\cite{yi} believed that insights can be ``sources or stimuli of other insights'' (e.g., insightful questions that an analyst had not thought about). Stasko~\cite{value} commented that the effects of knowledge on an analyst's mental model (e.g., learning a domain and confirming a hypothesis) are also insights.

Our work investigates data insights from the first perspective. During the interviews, we asked interviewees to recall insightful findings. We summarize the characteristics of these data insights and discuss their implications for designing tools that aim to automatically communicate data insights to users.

\subsection{Tools That Aim to Automate Data Insights}

{Existing tools that aim to automate data insights offer guidance~\cite{guidance} by suggesting potentially relevant data observations.} They often do so by recommending charts that reveal potentially interesting data patterns and/or providing textual descriptions of statistical data facts.

Many researchers have investigated tools that recommend noteworthy charts (e.g., a scatterplot with a high correlation or a line chart with an upward temporal trend) (e.g.,~\cite{datasite, dive, pilot, vizdeck}). These tools are often referred to as data-based recommendation systems~\cite{doris} or data query recommenders~\cite{general}. For instance, Foresight recommends “visual insights” based on metrics such as skewness and correlation~\cite{foresight}. SeeDB identifies bar charts that show a deviation from a reference~\cite{seedb}. Different from typical visualization recommendation systems (e.g., Voyager~\cite{voyager1} and Show Me~\cite{showMe}) that recommend perceptually-effective charts to users, these tools proactively identify trends and patterns in data based on the statistical properties of data and visualize these noteworthy trends and patterns as charts. 

In parallel, some researchers have developed tools that communicate statistical facts about data using textual descriptions (e.g., ``US cars have a higher average horsepower than Japanese cars'' for a dataset of cars). They often refer to such textual descriptions about data as data facts~\cite{voder, datashot}. For instance, Voder generates textual descriptions of charts and enables users to interact with the text to facilitate interpretation~\cite{voder}. TSI highlights prominent regions in temporal visualizations through automatic annotation~\cite{tsi}.

Such functionality has also emerged in commercial visualization platforms. For instance, Quick Insights in Power BI automatically identifies prominent data patterns~\cite{quickInsights}. Explain Data in Tableau proposes data-driven explanations for an outlying value~\cite{explainData}.

While some researchers call these automatically generated visualizations and/or textual descriptions ``insights''~\cite{algo1, foresight}, others feel that these recommendations, in their current state, are qualitatively different from data insights that are ``deep'' and ``complex''~\cite{stasko}. We recognize that what data insights are and how to generate charts and textual descriptions that users truly find insightful are still (and will still be) an ongoing conversation within the visualization community. We intend to contribute to the conversation by studying the characteristics of data findings visualization users consider insightful.

\section{Methodology}

\noindent \textbf{Participants}. We interviewed 23 practitioners (14 male, 9 female) from 19 organizations. Interviewees worked in 12 job sectors including consulting, retail, and education. The organizations ranged from solo entrepreneur, to start-up companies with less than 10 people, to large corporations with more than 100,000 people. The locations of the organizations spanned 5 US states (Georgia, Massachusetts, North Carolina, Pennsylvania, and Tennessee). We provide the detailed demographics of interviewees as a supplementary material.

As an inclusion criterion, interviewees were required to employ end-user visualization platforms in their jobs. They utilized a wide variety of visualization platforms including, among others, Tableau (23/23 interviewees), Power BI (10/23), Qlik (4/23), and Cognos (4/23). They had 2--20 years of experience with these systems.

We recruited interviewees through multiple channels. An author went to a Power BI user group meeting in Alpharetta and four Tableau user group meetings in Atlanta, Charlotte, and Philadelphia. We approached attendees in order to collect contact information to schedule interviews. We also emailed contacts in our professional networks and potential interviewees found through websites such as LinkedIn. Several interviewees helped identify colleagues and friends in their networks who they thought might be eligible for the study. Interviewees were not compensated.

\vspace{1mm} \noindent \textbf{Interviews}. An author conducted semi-structured interviews with the 23 practitioners between November 2019 and February 2020. The interviews were one-on-one except an interview with 2 interviewees. We interviewed the practitioners in their office (1/23), in researchers' lab (2/23), and via video-conferencing software (20/23). The interviews lasted between 40 and 75 minutes. With the permission of interviewees, we recorded the audio for subsequent analysis.

During the interviews, we probed into three topics: interviewees’ data analysis workflow, experiences with data insights, and initial perceptions of prototypes that automatically generate data facts. For the scope of this paper, we report on interviewees’ experiences with data insights. To understand this topic, we asked interviewees to describe a finding from their data they considered to be insightful and the characteristics of the finding that made it insightful.

\vspace{1mm} \noindent \textbf{Analysis}. An author manually transcribed the audios recorded during the interviews. The author segmented the transcripts into passages (most of them are single sentences) and applied open coding. He coded the transcripts throughout the interview study. 

During the coding process, the coder followed constant comparison and theoretical sampling in grounded theory~\cite{gt}. When coding a passage, he returned to the data repeatedly to compare it with other passages (constant comparison). He labelled related passages as the same category and grouped related categories into a dimension. For each dimension, he considered other possible categories that had not yet emerged in the coded data and focused on these categories in subsequent interviews and coding (theoretical sampling). The coder iteratively refined the categories and dimensions through frequent discussions with other researchers in the team.

\section{Who Are These Visualization Users?}

Before presenting the characteristics of the data insights mentioned by interviewees, we first describe their data analysis workflow to provide context for our findings. 

Reporting was a central activity in the workflow of interviewees. All interviewees employed end-user visualization platforms for crafting reports. Besides traditional reports (e.g., PDFs and presentation slides), most interviewees mentioned creating visualization dashboards (22/23). Audiences of the reports varied. They could be internal audiences in other departments (14/23) and clients (4/23). Interviewees also mentioned creating dashboards for audiences external to the organizations who were not clients (6/23). For example, institutional research departments at colleges created dashboards for prospective students and parents (2/23).

Interviewees played dual roles in the process of report creation: data analyst and visualization designer. 

As an analyst, interviewees conducted the analysis required for creating reports (23/23). Analysis for creating reports often involved measuring key performance indicators of some processes (15/23). Before creating reports, some interviewees conducted exploratory analysis to get familiar with the data (10/23) or look for interesting patterns (10/23). Spreadsheet applications such as Excel were the \textit{``go-to tool''} for such initial analysis (15/23). Interviewees often used Excel for inspecting tables (8/23) and creating pivot tables (7/23). When reporting statistical analysis, interviewees conducted the analysis using statistical software such as R and SPSS (3/23).

As a designer, interviewees designed dashboards for their audience (22/23). They often employed human-centered design~\cite{norman} for dashboard development (16/23). During the design of a dashboard, they first gathered requirements from end users. Some mentioned receiving a report specification from the end users (7/23). Design requirements might also be a product of discussions between interviewees and the end users (13/23). Throughout the design process, interviewees often demonstrated dashboard designs to the end users for feedback and iteratively refined the design accordingly (15/23). 

\section{Characteristics of Data Insights}

The qualitative analysis resulted in seven characteristics of data insights. Note that not all data insights have all the characteristics.

\vspace{1mm} \noindent \textbf{Actionable}. The analysis scenarios most interviewees described were closely tied to decision making (17/23). Many interviewees considered insights to be findings that were actionable (9/23).

Some interviewees described insights that enticed people to take actions (6/23). In I1's words, \textit{``I feel like anything that is an insightful finding has to have a call to action.''} I18, a marketing consultant, gave an example: \textit{``After we take a look at that data, we might find that you're spending completely like way too much money on print, way too much money on TV ads, and they're just not effective.''} 

Some insights informed what actions to take after interviewees learned about a need for actions (3/23). I16, a sales analyst, recalled an example where a data insight helped inform pricing strategies: \textit{``Seeing how was the margin on that product in California vs in Florida [...] and trying to figure out why one product is stronger in a location than in another [...] It would help us determine pricing if we knew that it's a competitive disadvantage.''} 

\vspace{1mm} \noindent \textbf{Collaboratively refined}. Multiple interviewees described scenarios where they conducted analysis or created dashboards for subject matter experts (12/23). While the experts had significant knowledge about their domains, they often lacked the expertise or time for the analysis (8/23). In contrast, visualizations users, while having the skills for the analysis, often did not possess complete domain knowledge (11/23). I7, a data analyst at a college, said, \textit{``As a data expert, I don't necessarily understand their data. I know how it's structured and I know what the field names are [...] but I don't necessarily always understand exactly how that data was created or the business process where they were created.''} 

As interviewees might lack the domain knowledge needed for an analysis, they often worked with the subject matter experts to understand whether a finding is insightful (7/23). I5, who worked at an institutional research department of a college, depicted how she reached out to the experts when finding something that appeared to be useful: \textit{``We have to go back to our med school advisor and ask is this something that is worth looking at [...] is there something that you want to see the output of.''}

\vspace{1mm} \noindent \textbf{Unexpected}. Interviewees also described insights that diverged from some expectations (7/23) with I19 stating, \textit{``I expected there to be some sort of relationship that was obvious, but I didn't see a relationship at all [...] It was unexpected. It's not what I thought it would be [...] So, that was insightful to me.''} I1 recalled a scenario where her clients stumbled upon an unexpected insight during data exploration: \textit{``I'm thinking of a company I've worked with [...] They just had so much equipment they didn't know how it was being used [...] Generally speaking, that equipment should just be used during business hours and I think when we started analyzing the data and saw that [...] people using equipment on the weekends and at night during off-business hours they're like oh my gosh what's going on.''} 

\vspace{1mm} \noindent \textbf{Confirmatory}. Interviewees mentioned scenarios where they had an \textit{``expectation,''} a \textit{``hunch,''} an \textit{``intuition,''} a \textit{``hypothesis,''} and a \textit{``mental model,''} and an insight occurred when expectations were confirmed (6/23). I11 provided an example: \textit{``When I was working for the state government agency, the project was to analyze traffic patterns [...] It may take an average car ten minutes to drive that whole length but we saw that it took a car two minutes [...] Once I saw that pattern, then I was able to build a visualization based upon my observation, and I found out later on doing some mathematical models that my eureka moment was correct.''} 

I15 and I5 believed that data insights that confirmed with some expectations happened more frequently than those that deviated from the expectations. I15 said, \textit{``I would say most insights that people generally are asked to present are more so data-driven reassurance because if you have someone that's really familiar with a process, they deal with it every day then more times than not what they're going to be shown is like I understand that, I'm not surprised.''} 

\vspace{1mm} \noindent \textbf{Spontaneous}. Chang et al.~\cite{chang} used ``spontaneous insights'' to refer to insights that occur when ``a problem solver suddenly moves from a state of not knowing how to solve a problem to a state of knowing how to solve it.'' Interviewees recalled similar spontaneous insights where with visualizations, they were able to \textit{``see data in a form that they haven't seen before''} (I3) and \textit{``answer or ask questions of the information that they weren't previously able to''} (I3) (3/23). 

Such spontaneous insights were often ascribed to the power of visualizations: Without visualizations, it was hard to gain insights into raw numbers, but with visualizations, the data made sense for the first time. I2, a consultant, recalled an interaction with some clients: \textit{``I am building a dashboard for a large health system [...] They [the clients] have never been able to see not only where they are, how the public thinks about them but also what the rest of the other competitors look like [...] So, [with the dashboard] there's a sense of getting pretty excited of being able to see not just kind of where are we but where's everybody else are at the same time.''} 

\vspace{1mm} \noindent \textbf{Trustworthy}. Several interviewees emphasized the importance of validation upon finding an insight (5/23). I4 worked with a professor on a public policy project and said, \textit{``This [the finding] actually made her [the professor] believe she could challenge one of MIT's recent publications which I didn't know if she actually did. I suggested her to really go back and check the data source [...] We want to make sure every step is accurate. We don't want to roughly publish something say hey MIT you are wrong and end up being maybe one of our student assistants or I made some mistakes in the process.''} 

Being able to trust data insights was crucial often because interviewees needed to communicate and report the insights (5/23). I13 noted that skepticism about insights was common in presentations: \textit{``There's a lot of issues of trust in the room [...] The mathematicians in the room they're going to want to know what was excluded, what mathematical principle was utilized, how did you get there, did you filter anything out, right or wrong if you filter anything out.''} 

Some interviewees were wary of reporting data insights without enough confidence in the insights (2/23). I22 said, \textit{``I'm definitely conscious all the time of accuracy and [...] I'm not going to put anything out there if I don't have this understanding and trust.''} In contrast, some expressed confidence in reporting when the insights were validated (2/23): \textit{``Once I get through the validation [of the findings], you can't tell me nothing about my report. I know my report is right. I know the work that I put in to test it''} (I10). 

\vspace{1mm} \noindent \textbf{Interconnecting}. Some interviewees believed that insights did not occur just by seeing a piece of information, but instead happened when a piece of information paired with domain knowledge or other contextual information (3/23): \textit{``There are findings as facts which are relatively not contestable. They are an association of math output of an equation [...] a known fact. An insight is when you put multiple pieces of this together in the context of a question''} (I19).

I19, a political science researcher, described how such connections happened: \textit{``I was looking at the relationship between US and Chinese foreign aids to African countries [...] I saw that there's absolutely no relationship between where China used money and where US used money [...] It's like there's no competition which was really interesting to me [...] It made a connection to some other things in my head about how US gives aids and the US gives mostly like relief aids after a disaster [...] and they don't give any infrastructure aid where China gives countries aids to invest in infrastructure projects [...] So a long way to say the insight was based on the project. It was connected to other things I am thinking about necessarily.''} 

In essence, an observation from data often makes little sense when it stands on its own. Yet, other information provides context for the observation and brings new meanings to the observation. Insights occur when different information sources coalesce.

I22 said that the other information that provided context for a data observation could come from other people in a meeting: \textit{``We've seen weird blips with check average just suddenly falls [...] and then someone has information that there's a promotion and we're giving away some products, and it's stuff like that is I find really insightful.''} 

\section{Discussion}

For a car data set, existing systems that aim to automate data insights may recommend a bar chart along with a textual description such as “US cars have a higher average horsepower than Japanese cars.” Yet, our results bolstered prior work (e.g.,~\cite{north, chang}) that explicates the complexity of data insights: Data insights represent nuanced understanding of the data shaped by a user’s mental model as well as the social and organizational context where findings are discovered. The vision that automated systems can extract findings users consider insightful will require system designers to think beyond purely mining interesting visualizations and communicating textual descriptions of data facts. While understanding what data insights are and how to develop systems that can truly automate data insights will be an ongoing challenge, we hope to take a step in articulating some considerations in designing tools for automatically generating charts and data facts that are closer to the holy grail of data insights.

\subsection{Designing ``Automated Insight'' Tools}

\noindent \textbf{Information sources}. Interviewees often believed that data insights were \textit{interconnecting}: Arriving at data insights involves gathering information through data analysis (e.g., by inspecting and manipulating a visualization), connecting the information with domain knowledge and other contextual information, and applying intuition to piece different sources of information together. To provide similar data insights, techniques could be developed to algorithmically draw connections among statistical facts in data, domain knowledge, and contextual information. Accomplishing this goal, however, entails tackling several research challenges. First, domain knowledge and contextual information are often not in the data. Second, even if we have ways to represent and retrieve domain knowledge and contextual information, drawing intuitive connections among information like human intuition does is technically challenging and might require solutions beyond a simple rule-based approach.

\vspace{1mm} \noindent \textbf{Trust}. Some interviewees felt that data insights should be \textit{trustworthy}. Interviewees who believed in the importance of being able to trust the insights from data often went through rigorous validation of the insights to ensure their correctness. To many interviewees, developing trust in data insights requires working through the analysis manually and verifying the validity of the insights---there is a process involved. In contrast, tools that automate the production of data insights abbreviate the process of manual analysis. Without going through the process of validating automatically generated visualizations or data facts, users may find these recommendations difficult to trust. To inspire user trust in the recommendations, it can be a good idea to improve transparency in the automation. For example, an automated system could help users understand how the visualizations and data facts are generated. Validation mechanisms could also be incorporated into such systems to enable users to verify the correctness of the recommended data observations.

\vspace{1mm} \noindent \textbf{Mental Model}. Many characteristics of data insights (\textit{unexpected}, \textit{confirmatory}, and \textit{spontaneous}) we observed were tied to a user’s mental model. As users extract facts from data and interpret the facts through a mental model, an insight might occur when the facts deviate from the mental model, align with the mental model, or somehow trigger a light-bulb moment. Many existing tools that aim to automatically provide data insights often define some interestingness functions for ranking charts or data facts. For instance, some systems rank scatterplots based on correlation coefficient (e.g.,~\cite{foresight}). The assumption is that users often find a highly-correlated scatterplot more insightful. However, this assumption may not always hold true. An alternative design idea is to directly acquire users' mental models (e.g., users' expectations about the data). For instance, Choi et al.~\cite{concept} proposed concept-driven visual analytics. They envisioned a system that enables users to externalize their expectations about the data through natural language (e.g., I expect the US to have the highest GDP per capita in the world). The system will then recommend relevant visualizations to users (e.g., showing a bar chart that reveals the GDP per capita of different countries with a textual description that indicates the true rank of the US). Richer information about a user’s mental model could increase the likelihood that users consider the recommended visualizations and data facts to be insightful.

\vspace{1mm} \noindent \textbf{Context}. Our results also revealed that data insights were dependent on the social and organizational context: Interviewees often believed that insights should help inform decisions (\textit{actionable}) because they used data for decision making; they often collaborated with subject matter experts to refine a data insight (\textit{collaboratively refined}) since they might lack the domain knowledge for conducting the analysis. Many existing tools such as Quick Insights (Fig.~\ref{powerBI}) present general-purpose charts and data facts without considering users' domains and work context. To increase the likelihood that users consider the recommended charts and data facts insightful, “automated insight” tools could be tailored to more specific use cases. This implies a need for system designers to identify what findings are insightful and actionable to a specific user group during the design process, as opposed to relying solely on designer intuitions of what insight is.

\subsection{Connection to Existing Definitions}

Some interviewees considered data insights to be findings that deviated from some expectations (\textit{unexpected}) and triggered a light-bulb moment (\textit{spontaneous}). The two characteristics were similarly identified by North~\cite{north} and Chang et al.~\cite{chang} respectively. Interviewees often believed that multiple sources of information coalesced to form data insights (\textit{interconnecting}). This echoes North’s definition that insight is complex (``involving a large amount of data in a synergistic way'')~\cite{north}. Comments regarding data insights being \textit{collaboratively-refined} and \textit{actionable} indicated interviewees’ desire to glean insights that were relevant to the goals of their organizations. These two characteristics clarify what North~\cite{north} means by ``relevant'' insights to our interviewees. {Furthermore, Sacha et al.~\cite{trust} similarly highlight the importance of building trust in data findings during insight generation (\textit{trustworthy}).} Insights being \textit{confirmatory}, however, appear to be less emphasized in the literature. Our interviews could provide an empirical basis for confirming the prevailing definitions of insight in the visualization community while identifying new perspectives for considering what data insights are.

\subsection{Limitations}

Our study suffered from the same limitations as typical interview studies. First, interviewees were asked to recall past experiences. Their depictions might be imprecise due to a limited ability to recall the past. Moreover, the number of interviewees who mentioned a view or an activity does not reflect the frequency of the view or activity among the whole population of visualization users. Surveys are better suited for quantifying a phenomenon. We also recognize that the characteristics of data insights we identified may not be complete. Additional interviews with a broader set of visualization users could augment our findings. {Finally, Tableau users might be overrepresented in our study. In future studies, researchers could recruit a more representative set of interviewees based on surveys of visual analytics systems (e.g.,~\cite{survey2}) and their market shares.}

\section{Conclusion}

We have reported on seven characteristics of data insights observed from an interview study with 23 visualization practitioners. Based on the findings, we distilled considerations for designing automated systems that communicate data insights to users. While articulating what insight is and how to develop tools that automate data insights will still be an ongoing challenge, we have taken a step in unpacking some of the nuances of insights and ways researchers could adopt to move closer to the ultimate vision of an ``automated insight'' tool.

\bibliographystyle{abbrv-doi}

\bibliography{template}

\onecolumn

\section*{Appendix: Interviewees' Demographics}

\begin{figure*}[h]
	\centering
	\includegraphics[width=\linewidth]{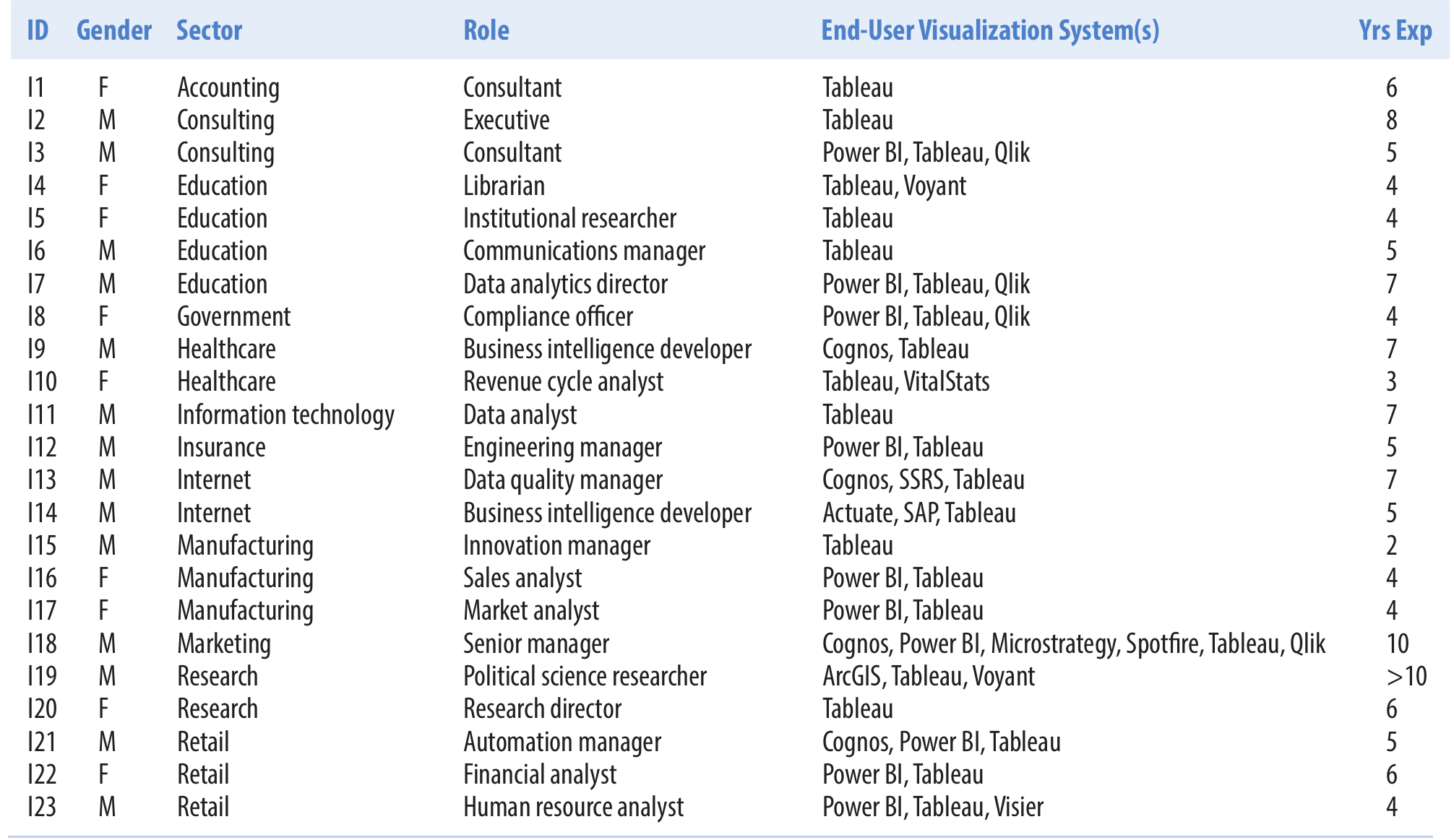}
	\caption{Demographics of interviewees. The table shows their job sectors, roles, end-user visualization systems used at work, and years of experience with these systems. Interviewees are sorted by job sector.}
\end{figure*}

\end{document}